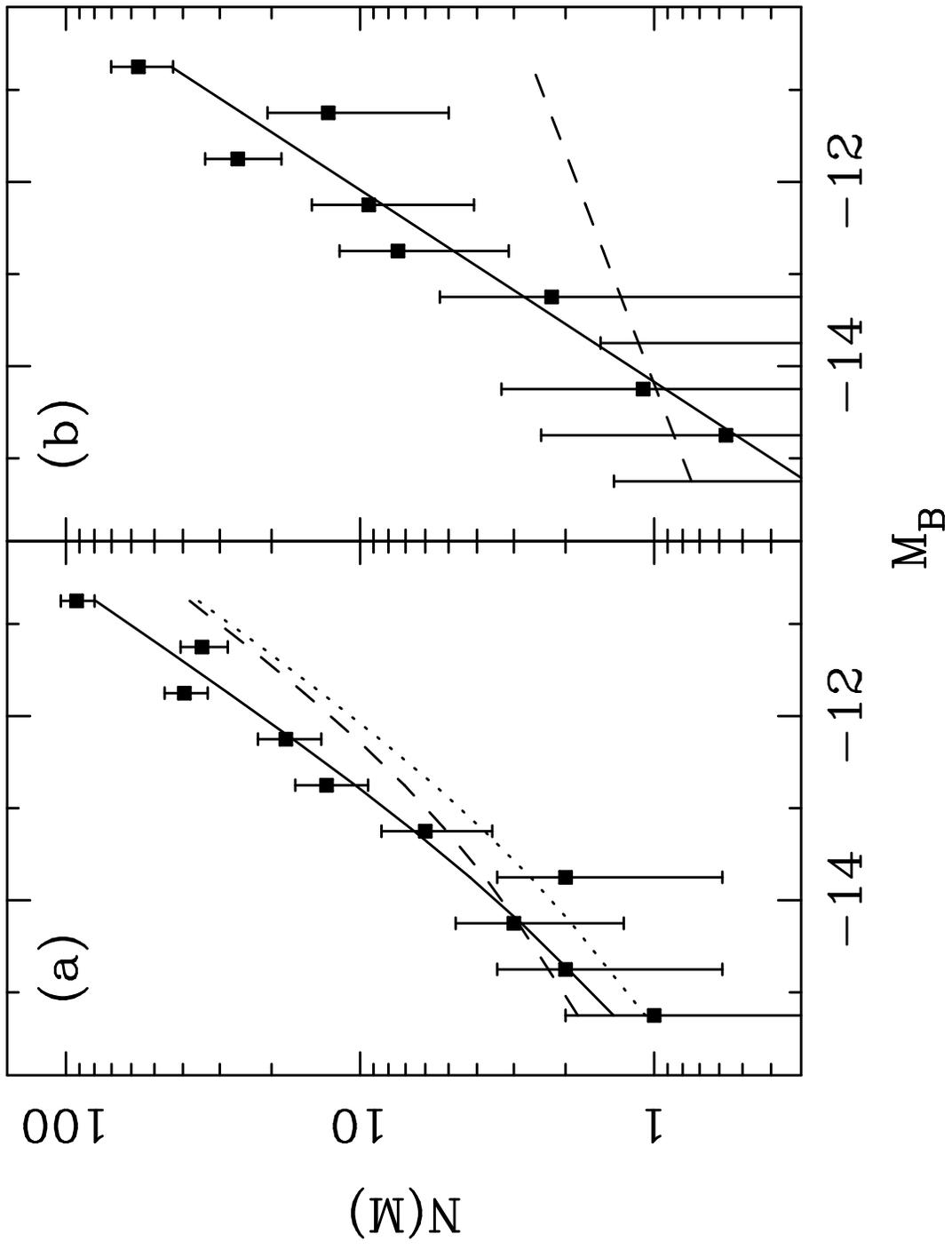

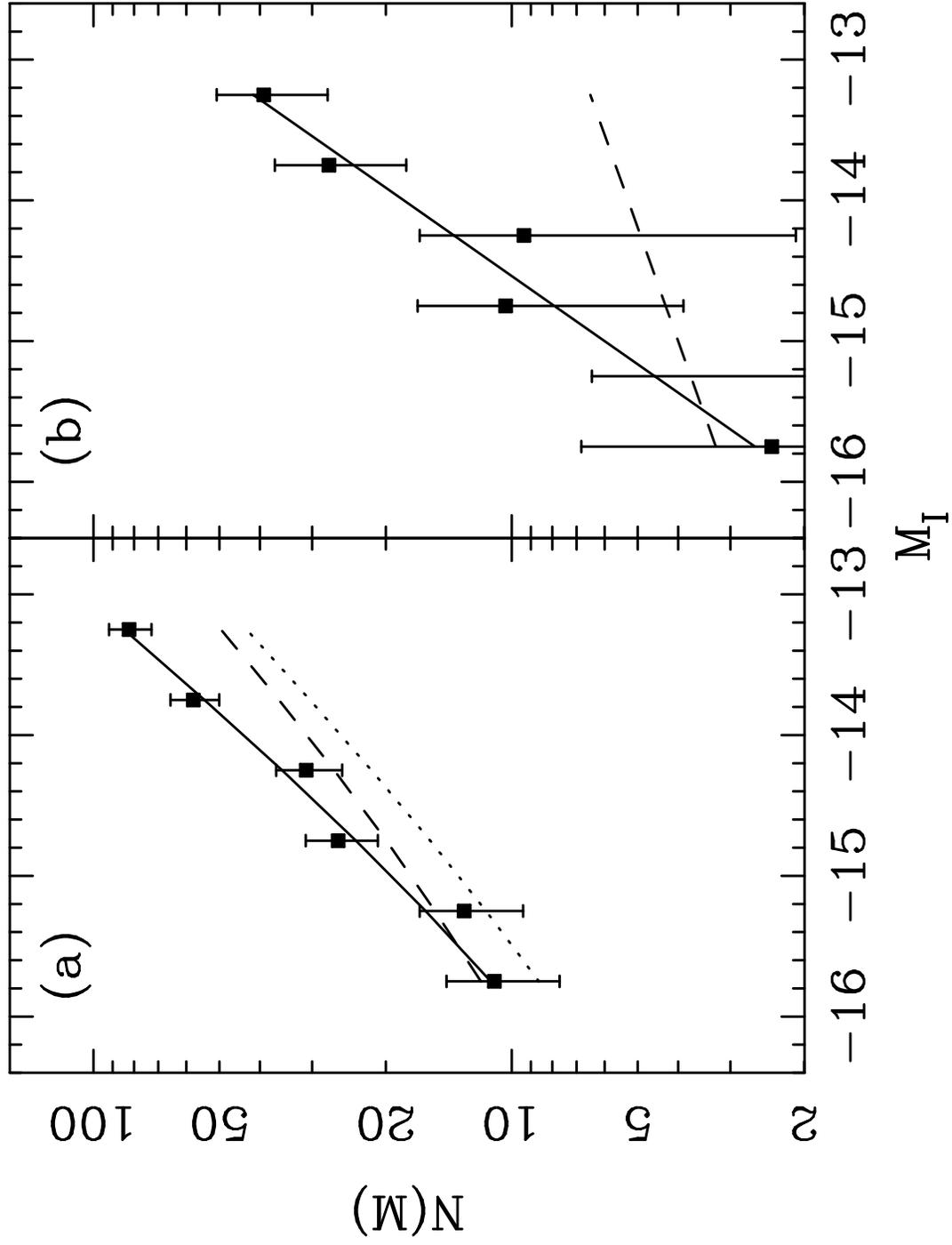

# Evidence for Steep Luminosity Functions in Clusters of Galaxies[2]

Roberto De Propris

Department of Physics and Astronomy, University of Victoria, PO Box 3055, MS 7700, Victoria, BC, Canada, V8W 3P6

e-mail: propris@otter.phys.uvic.ca

Christopher J. Pritchet[1]

Department of Physics and Astronomy, University of Victoria, PO Box 3055, MS 7700, Victoria, BC, Canada, V8W 3P6

e-mail: pritchet@clam.phys.uvic.ca

William E. Harris[1]

Department of Physics, McMaster University, Hamilton, ON, Canada, L8S 4M1

e-mail:harris@physun.physics.mcmaster.ca

Robert D. McClure[1]

Dominion Astrophysical Observatory, Herzberg Institute of Astrophysics, National Research Council, 5071 W. Saanich Road, Victoria, BC, Canada, V8X 4M6

e-mail: mcclure@dao.nrc.ca

---

# ABSTRACT

Luminosity Functions have been obtained for very faint dwarf galaxies in the cores of four rich clusters of galaxies (Abell 2052, Abell 2107, Abell 2199 and Abell 2666). It is found that the luminosity function of dwarf galaxies rises very steeply in these clusters, with a power-law slope of $\alpha \sim -2.2$ (down to limiting absolute magnitudes $M_I \sim -13$ and $M_B \sim -11$ for $H_0 = 75$ km s$^{-1}$ Mpc$^{-1}$). A steepening of the luminosity function towards low luminosities may in fact be a common feature of both cluster and field populations. Such a result may explain the observed excess counts of faint, intermediate redshift galaxies in the Universe, without resorting to more exotic phenomena. An alternate explanation is that star formation in dwarf galaxies is less affected by gas loss in the richest clusters, because of the dense, hot intracluster medium found in such environments.

## 1. Introduction

The Luminosity Function (LF) of faint galaxies is of importance for many reasons. Dwarf galaxy numbers depend on the properties of the primordial power spectrum of density fluctuations; for example, the currently fashionable "cold dark matter" (CDM) scenario predicts a fluctuation power spectrum tilted towards low mass fluctuations (e.g. review by Ostriker 1993), and simulations of galaxy evolution in a CDM-dominated Universe show, not surprisingly, a large excess of dwarf galaxies (White & Frenk 1991). Furthermore, dwarf galaxies are fragile entities; their subsequent evolution, not to mention survival, is strongly affected by a variety of galaxy evolution processes, as reviewed by Ferguson & Binggeli (1994). Finally, a steep dwarf galaxy LF may be required to explain the large numbers of galaxies seen at intermediate redshift (Marzke *et al.* 1994a,b; Koo *et al.* 1993) if more exotic explanations are to be avoided. It is therefore of great current interest to know the shape of the dwarf galaxy LF.

In contrast to the bright galaxy LF, which is comparatively well established (Binggeli *et al.* 1988), the faint end of the LF ($M_B \geq -16$) is poorly known either in clusters or in the field, because of the great difficulty in obtaining complete samples of faint, low surface brightness objects. Similarly, compact galaxies (e.g. Zwicky compacts) are also missed as they are confused with stars. Dwarf galaxies should, on the other hand, be sufficiently numerous in the cores of *rich* clusters of galaxies that an accurate dwarf galaxy LF can be obtained. The problem of contamination by foreground and background objects can be handled statistically if the core density of the cluster is high enough. Furthermore, the fact that the nearest rich clusters of galaxies are somewhat more distant ($\geq 5\times$) than Virgo and Fornax may actually be an advantage for dwarf galaxy completeness: low surface brightness galaxies tend to become easier to detect as their angular size decreases (i.e. distance increases).

In this paper we report on new observations of the dwarf galaxy luminosity function down to very faint magnitudes. These observations demonstrate that the faint galaxy LF in rich clusters is very steep, in contrast to observations at brighter magnitudes.

## 2. Observations and Data Reduction

The data used for this project originated in earlier observations (Pritchet & Harris 1990; Harris *et al.* 1995) using the Canada-France-Hawaii 3.6m Telescope. The cluster Abell 2199 (centred on its associated cD galaxy NGC 6166) was observed during the nights of 1988 May 18-19 (UT) in the $B$ band, using a RCA $640 \times 1024$ CCD. Observations of Abell 2052 (centred on its associated cD galaxy UGC 9799), A2107 (UGC 9958) and A2666 (NGC 7768) were taken in the $I$ band during the nights of 1990, June 26-30, using a SAIC $1024 \times 1024$ CCD with the High Resolution Camera (HRCam — McClure *et al.* 1991). Exposure times, redshifts and field sizes are given in Table 1 for all clusters and for one $I$ band non-cluster field used to monitor foreground and background contaminants. Details of the observations, initial data reduction, and preprocessing can be found in Pritchet & Harris (1990) for A2199, and in Harris *et al.* (1995) for A2052, A2107 and A2666.

All images were then analyzed in the following way: *(i)* Isophotes were computed for the central galaxy in each frame. These isophotes were used to produce a smoothed model of the central galaxy; this model was subtracted off each frame. *(ii)* Each image was then convolved with a lowered Gaussian, having a $FWHM$ approximately equivalent to the stellar $FWHM$ and with a $4 \times FWHM$ kernel. Initially, all objects above a $4\sigma$ deviation from the mean sky level in the convolved frame were flagged. For most of the objects considered in this study the choice of detection threshold is unimportant. *(iii)* For each object we computed image moments (Kron 1980) $r_1$ and $r_{-2}$ and a Petrosian radius $r_p$ equivalent to the radius for maximum signal to noise ratio (or, equivalently, max $I(<r)/r$). From large numbers of tests $r_1$ is almost identical to $r_p$; $r_1$ is itself defined by integration of the profile of each object down to 0.003 of the sky level, at which point the integration is stopped. *(iv)* All objects having $r_1 \leq 0$ were discounted as spurious; photometry was obtained for all remaining objects in an aperture of radius $2r_p$. This procedure turns out to give almost identical results to the Kron (1980) photmetry algorithm, but is somewhat more robust. *(v)* Artificial objects having $FWHM$ approximately equal to the median $FWHM$ of all objects in our frames were then introduced in our images and steps 3 - 5 were repeated, to produce completeness estimated for our frames.

We obtained number counts to about 1 mag brighter than the brightest globular cluesters in these galaxies (Pritchet & Harris 1990, Harris *et al.* 1995). Tables 2 and 3 below show that we are nearly complete at all magnitudes, and that completeness corrections are not very large (although they have been included in the subsequent analysis).

Because of the relatively modest seeing ($\sim 0''.8 - 1''.0$) we were unable to safely discriminate between stars and galaxies at the magnitudes of interest. Compact dwarfs may also masquerade as stars, especially at the distance of our clusters. We have therefore decided to handle contamination by stars and galaxies statistically. Field contamination by stars was estimated from the starcount model of Bahcall & Soneira (1980, 1981). Contamination from faint background galaxies was accounted for in the following way. For the $I$ band, we used our background field (Table 1), removed the stars using the Bahcall & Soneira program, and fitted the counts in this frame with $\log N = a + bm$, with $b = 0.34$ (Lilly *et al.* 1991). The resultant counts are given by $N = 1.056 \times 10^{0.4(I-19)}$ (4.42 arcmin$^2$, 0.5 mag bins); this agrees very well with the counts of Tyson (1988), and also of Lilly *et al.* (1991) after correction from the $I_{AB}$ to $I$ system. For the $B$ band frame we used the average of galaxy counts from Tyson (1988), Lilly *et al.* (1991) and Metcalfe *et al.* (1991), to obtain $N = 0.253 \times 10^{0.45(B-20)}$ (6.54 arcmin$^2$, 0.5 mag bins — These sources agree to better than $\pm 15$ %). Tables 2 and 3 show the raw data, completeness corrections and completeness corrected data, background contamination (stars and galaxies) and final counts in each frame. The magnitudes quoted in these tables refer to the midpoint of each bin. Note that the errors quoted in these tables are "illustrative" only, as all of our subsequent analysis was carried out using general maximum likelihood methods assuming a Poisson probability distribution. As HRCam frames span only about 3.1–3.7 arcmin$^2$ we decided to sum all of our $I$ band data to improve our statistics. From Tables 2 and 3, a number of faint galaxies can be seen to be present in A2052, A2107, and A2199; the effect is not as pronounced in A2666. Using maximum likelihood techniques the data were fitted with an equation of the form:

$$N_{obs}(m) = f(m)[s(m) + g(m) + K\, 10^{-0.4m(\alpha+1)}], \qquad (1)$$

where $s(m)$ and $g(m)$ are the (assumed) contributions of foreground stars and background galaxies, $f(m)$ is the known incompleteness fraction, and $K$ and $\alpha$ are determined by the program. For the $B$ band data we find $\alpha = -2.16 \pm 0.18$ and for the $I$ band data (A2052 + A2107 + A2666) we find $\alpha = -2.28 \pm 0.30$. Nearly identical results are found with nonlinear weighted least squares, but of course maximum likelihood techniques (assuming Poisson probabilities) are preferred given the small numbers of objects. The quoted errors in $\alpha$ correspond to $\pm 1\sigma$. We can formally rule out $\alpha$ shallower than $-1.5$ at the $> 99$ % level; $\alpha$ shallower than $-1.7$ can be ruled out at the 99% (97%) level for the $B$ ($I$) data. These results are quite insensitive to the range of data used in the fit, or to whether or not the data is completeness corrected.

## 3. Discussion

The slope of the background counts is similar (within errors) to the slope of our LFs and this raises the possibility that an incorrect subtraction of background is responsible for our result. Lowering the number of contaminants has little effect on $\alpha$, but increases the significance that $\alpha \simeq -2$. The effect of raising the number of contaminants is obtained from the following argument. The field-to-field variance in number counts can be calculated from the angular correlation function of faint galaxies (Peebles 1975, 1980). Taking faint galaxy correlation amplitudes of Pritchet & Infante (1992), and field sizes from table 1, it is found that the probability that the background counts exceed the expected values by more than 1.4 is $< 1\%$. Now, raising the number of contaminants by 40% still produces $\alpha < -1.5$ at the $> 99.5\%$ confidence level in the $B$ data; for the $I$ data the significance of the result is not as great but we can still rule out a slope of $\alpha = -1.3$ with $> 90\%$ confidence. However, such a high level of background is *extreme* and it is very unlikely that it would be present for *four* widely separated clusters.

Figures 1 and 2 show the data and the best fit to the sum of cluster dwarf galaxies and contaminants. The dashed line represents a LF with a slope of $\alpha = -1.3$ forced through one of the brighter bins, and illustrates that there is a considerable excess of faint galaxies over a "Virgo" LF.

In order to verify the robustness of our result and its stability for different normalizations of the background counts, we have conducted a series of Monte Carlo simulations using our maximum likelihood estimator. The procedure for each Monte Carlo realization was as follows. *(i)* $n_c$ cluster galaxies were chosen from a LF with $\alpha = -2$. Here $n_c$ is drawn from a Gaussian distribution with mean $N_c^0$ and dispersion $\sqrt{N_c^0}$, where $N_c^0$ is the (completeness-corrected) number of galaxies (sum of column 6 in Tables 2 and 3). *(ii)* $n_g$ foreground/background objects were drawn from a LF as described in §2 above. The number of objects $n_g$ is drawn from a Gaussian distribution with mean $= N_g^0$ and dispersion $2\sqrt{N_g^0}$, where $N_g^0$ is the (completeness-corrected) number of contaminants (sum of column 5 in Tables 2 and 3 — The extra factor of 2 in the dispersion allows for the effect of clustering; see Table 5 of Metcalfe *et al.*, and also the discussion above.) *(iii)* Some of the objects in each bin were discarded to allow for the completeness fraction. *(iv)* Maximum likelihood techniques were used to fit the data to Eqn. (1).

The main result of thousands of simulations of this sort is that the input $\alpha$ is recovered, in the mean, to very high accuracy; furthermore the errors in $\alpha$ are as quoted earlier. Statistically, we thus find $> 99.5\ \%$ ($> 98\%$) confidence for the $B$ ($I$) data that $\alpha$ is steeper than $-1.5$ for dwarf galaxies in these rich clusters. It should again be noted that these simulations include the effect of stochastic variations in the background due to both Poisson fluctuations and also to galaxy clustering.

Another possibility is that our LF is being artificially steepened by photometric errors (the Eddington correction; — Kron 1980 and references therein). This is most relevant when photometry is carried out close to the frame limits, which we are well above, but it may affect some of our fainter bins, that contribute the most to our result. Simulations show that there is no evidence that an Eddington correction is needed and place a strong upper limit of $\sim 0.2$ on its magnitude, which would depress $\alpha$ to $-2.0$. It is therefore extremely unlikely that photometric errors are responsible for our result.

We have also considered the possibility that spurious detections may contribute to our counts. We regard this as unlikely, as all of our objects are well above the image thresholds (as witnessed by the small completeness corrections). Furthermore, we have visually examined every detected object in the CCD frames to make sure that it is real. Examination of these objects show that they are, on average, slightly resolved and slightly elongated. Their linear diameters, as determined from the $r_p$ parameter, are about 1 to 1.2 kpc, which is consistent with values for Local Group dwarfs (Lin & Faber 1983).

Errors in our completeness corrections are also unlikely to contribute to our results. If completeness has been overestimated, our corrections are small enough that they are not crucial for our conclusions, whereas if we had underestimated completeness, our result would become even more significant. Our previous work in these clusters reached much fainter limits ($I > 24.5$; $B > 25.5$), albeit for starlike objects and we are *well above* image thresholds for such objects. In any case our results are hardly affected if completeness corrections are ignored.

We cannot rule out, based on our lack of accurate structural and color information, a population of young, bright globular clusters as causing our excess of objects. On the other hand young globulars are generally found in the proximity of merging objects (e.g., NGC 1275 — Holtzman *et al.* 1993; NGC 7252 — Whitmore *et al.* 1992) and none of our central cD's shows evidence of a recent merger. Furthermore, most of the objects that we have detected do seem to be resolved (see also Binggeli *et al.* 1984 on this point). Nevertheless, it would be of interest to obtain HST images of our fields to obtain more quantitative estimates of the morphology of our dwarf galaxies.

The objects which have been found in this study therefore most likely represent a population of dwarf galaxies in the cores of our clusters. Our data suggest that a large population of galaxies is present at faint magnitudes, with a significantly steeper LF than the canonical $\alpha = -1$ to $-1.3$ Schechter function which is fitted at brighter magnitudes.

A comparison of this result with other determinations of $\alpha$ in the literature is somewhat confusing. The field surveys of Efstathiou *et al.* (1988) and Loveday *et al.* (1992) give $\alpha \simeq -1.0$. Cluster and group values seem somewhat higher. We have already noted the value of $\alpha = -1.3$ derived for the Virgo and Fornax clusters (Binggeli *et al.* 1985, Sandage *et al.* 1985, Ferguson & Sandage 1988); this value of $\alpha$ may steepen to $\sim -1.6$ with the addition of some low surface brightness dwarfs, coupled with a different analysis technique (Bothun *et al.* 1991). Ferguson & Sandage (1991) find similar results for other groups, with $\alpha$ generally in the range $-1.6 \leq \alpha \leq -1.3$. Tully (1988) advocates $\alpha = -1$ from an analysis of 6 nearby groups; but steeper values of $\alpha$ cannot be ruled is $L^*$ is altered. van den Bergh (1992) finds that $\alpha = -1$ fits the Local Group data quite well. But there are very few galaxies in the Local Group, and it is quite likely that incompleteness sets in in the Local Group data below absolute magnitude $M_B \simeq -12$; the Local Group results should therefore be viewed with some caution.

There appears to be growing evidence that, even if the LF at bright magnitudes is well-fitted by a Schechter function with $\alpha$ as low as $-1$, LF's for fainter galaxies may be steeper. The work of Marzke *et al.* (1994a,1994b) for field galaxies shows that a single Schechter function does *not* provide a good fit to the available data. Of more relevance to this paper, the Coma cluster (whose environment most resembles that of our rich clusters) possesses a LF that is not well fitted by a Schechter function (Thompson & Gregory 1993). The value of $\alpha = -1.4$ derived for this cluster at faint magnitudes (down to $M_B \sim -16$), differs from that closer to $L^*$. In fact, the LF for this cluster has recently been found to steepen at faint magnitudes, reaching values as steep as we find for our clusters (Bernstein *et al.* 1995). There exist some other data in the literature suggesting that the LF may have an upward inflection at faint magnitudes, in agreement with our result and the Bernstein *et al.* result. The LF derived from the CfA redshift survey catalog shows an excess over an $\alpha = -1.25$ LF at faint magnitudes (Marzke *et al.* 1994a, da Costa *et al.* 1994) and the Sd-Im LF has a slope of $\alpha = -1.9$ (Marzke *et al.* 1994b, although with larger errors). Driver *et al.* (1994) have derived a slope of $\alpha = -1.8$ for the distant cD cluster A963, although this is a difficult measurement because of the distance of the cluster ($z = 0.20$). The findings presented here reinforce these trends and suggest that the LF may steepen at faint magnitudes.

A steep LF for dwarf galaxies would have considerable importance for our understanding of faint galaxy number counts, if it is a general property of all dwarf galaxy populations. It is well known that faint number counts show a considerable excess over "no evolution" models (Tyson 1988, Lilly *et al.* 1991, Metcalfe *et al.* 1991) whereas redshift distributions are consistent with little or no evolution (Broadhurst *et al.* 1988, Cowie *et al.* 1991, Colless *et al.* 1990, 1993). Koo *et al.* (1993) have recently shown that this dilemma may be solved, without recurring to nonstandard models, if the LF is steep fainter than $M_B \simeq -15$ and if $\Omega_0$ is low. Our results appear to support such a scenario.

However, an alternate hypothesis (as opposed to a universally steep LF) is that the dwarf galaxy LF may be strongly affected by environmental factors. Babul & Rees (1992) have argued that dwarf galaxies may more easily survive (and convert a larger fraction of their initial baryonic mass into stars) near the centers of rich clusters; this is because gas pressure from the intracluster medium (ICM) prevents them from losing their gas to supernova-induced winds. Our clusters are all Bautz-Morgan I or II cD clusters and are therefore very gas rich. This may explain the steep LF that we find for dwarf galaxies in these clusters (recent evidence concerning the points above can be found in an RAS specialist discussion in *Observatory* vol. 114, p. 164ff). The Babul and Rees conjecture cannot be used (yet) to predict values of $\alpha$ in different environments, but is broadly consistent with a steepening of $\alpha$ values for dwarfs in denser, more gas rich environments. Although most of the data above is consistent with this hypothesis, it is very difficulty to understand how *field* Sd-Im galaxies could have a slope as steep as $\alpha = -1.9$ in this picture.

## 4. Conclusions

Analysis of deep CCD frames for the cores of rich Abell clusters has allowed us to produce luminosity functions with power-law slopes as steep as $d\log N(M)/dM \approx 0.5$ ($\alpha \sim -2.2$). This may be an universal characteristic of the galaxy LF at faint magnitudes or a product of environmental conditions in the cores of rich clusters. To decide between these two hypotheses will require further observations of clusters with a wider range of properties than initially chosen. Such observations are now underway.

We wish to thank Henrik Vedel and David Hartwick for helpful comments. We also wish to thank an anonymous referee for some very helpful remarks. The work of CJP and WEH is supported by grants from the Natural Sciences and Engineering Research Council of Canada. RDeP would like to thank the University of Victoria for a Graduate Fellowship.

TABLE 1
Summary of Observations

| Field | Redshift[a] | Exposure Time | Field Size[b] [arcmin$^2$] | FWHM |
|---|---|---|---|---|
| A2052 | 0.035 | $12 \times 1000$s | 3.53 | $0''.8$ |
| A2107 | 0.042 | $12 \times 1000$s | 3.68 | $0''.8$ |
| A2199 | 0.031 | $6 \times 1800$s | 6.54 | $1''.0$ |
| A2666 | 0.026 | $6 \times 1000$s | 3.13 | $0''.8$ |
| A2052bkg | — | $9 \times 1000$s | 4.42 | $0''.8$ |

[a] From Abell, Corwin, and Olowin (1989).

[b] Solid angle coverage after removing (i) vignetted areas due to HRCam optics, and (ii) areas near the center of the cD galaxy with I(galaxy) > I(sky).

TABLE 2
Observational Data – $I$ band

| I | $M_I$ | $N_{raw}$ | $N_{corr}$ | $N_{bkgd}$ | $N_{cluster}$ |
|---|---|---|---|---|---|
| | | | A2052 | | |
| 19.98 | −15.75 | 5 ± 2.24 | 5 ± 2.24 | 3.15 ± 1.77 | 1.85 ± 2.85 |
| 20.48 | −15.25 | 5 ± 2.24 | 5 ± 2.24 | 4.17 ± 2.04 | 0.83 ± 3.03 |
| 20.98 | −14.75 | 10 ± 3.16 | 10 ± 3.16 | 5.58 ± 2.36 | 4.42 ± 3.94 |
| 21.48 | −14.25 | 8 ± 2.83 | 8 ± 2.83 | 7.61 ± 2.76 | 0.39 ± 3.95 |
| 21.98 | −13.75 | 16 ± 4 | 16 ± 4 | 10.57 ± 3.25 | 5.43 ± 5.15 |
| 22.48 | −13.25 | 37 ± 6.08 | 39.36 ± 6.47 | 14.89 ± 3.86 | 24.47 ± 7.53 |
| | | | A2107 | | |
| 20.38 | −15.75 | 6 ± 2.45 | 6 ± 2.45 | 3.78 ± 1.94 | 2.22 ± 3.13 |
| 20.88 | −15.25 | 4 ± 2 | 4 ± 2 | 5.11 ± 2.26 | −1.11 ± 3.02 |
| 21.38 | −14.75 | 10 ± 3.16 | 10 ± 3.16 | 7.05 ± 2.66 | 2.95 ± 4.13 |
| 21.88 | −14.25 | 17 ± 4.12 | 17 ± 4.12 | 9.88 ± 3.14 | 7.12 ± 5.18 |
| 22.38 | −13.75 | 28 ± 5.29 | 29.79 ± 5.63 | 14.02 ± 3.74 | 15.77 ± 5.76 |
| 22.88 | −13.25 | 28 ± 5.29 | 34.15 ± 6.45 | 20.10 ± 4.48 | 14.05 ± 7.05 |
| | | | A2666 | | |
| 19.34 | −15.75 | 0 | 0 | 1.68 ± 1.30 | −1.68 ± 1.30 |
| 19.84 | −15.25 | 4 ± 2 | 4 ± 2 | 2.24 ± 1.50 | 1.76 ± 2.50 |
| 20.34 | −14.75 | 6 ± 2.45 | 6 ± 2.45 | 3.03 ± 1.74 | 2.97 ± 3.00 |
| 20.84 | −14.25 | 6 ± 2.45 | 6 ± 2.45 | 4.16 ± 2.04 | 1.84 ± 3.19 |
| 21.34 | −13.75 | 12 ± 3.46 | 12 ± 3.46 | 5.78 ± 2.40 | 6.22 ± 4.21 |
| 21.84 | −13.25 | 9 ± 3 | 9 ± 3 | 8.07 ± 2.84 | 0.93 ± 4.13 |

TABLE 3
Observational Data – $B$ band

| B | M $_B$ | N $_{raw}$ | N $_{corr}$ | N $_{bkgd}$ | N $_{cluster}$ |
|---|---|---|---|---|---|
| | | | A2199 | | |
| 20.21 | −15.25 | 1 ± 1 | 1 ± 1 | 1.07 ± 1.03 | −0.07 ± 1.44 |
| 20.71 | −14.75 | 2 ± 1.41 | 2 ± 1.41 | 1.43 ± 1.20 | 0.57 ± 1.85 |
| 21.21 | −14.25 | 3 ± 1.73 | 3 ± 1.73 | 1.91 ± 1.38 | 1.09 ± 2.22 |
| 21.71 | −13.75 | 2 ± 1.41 | 2 ± 1.41 | 2.63 ± 1.62 | −0.63 ± 2.15 |
| 22.21 | −13.25 | 6 ± 2.45 | 6 ± 2.45 | 3.77 ± 1.94 | 2.23 ± 3.13 |
| 22.71 | −12.75 | 13 ± 3.61 | 13 ± 3.61 | 5.57 ± 2.36 | 7.43 ± 4.31 |
| 23.21 | −12.25 | 17 ± 4.12 | 17.89 ± 4.34 | 8.56 ± 2.93 | 9.33 ± 5.24 |
| 23.71 | −11.75 | 36 ± 6 | 39.56 ± 6.59 | 13.50 ± 3.67 | 26.06 ± 7.54 |
| 24.21 | −11.25 | 30 ± 5.48 | 34.48 ± 6.30 | 21.66 ± 4.65 | 12.82 ± 7.83 |
| 24.71 | −10.75 | 58 ± 7.62 | 82.86 ± 10.88 | 35.36 ± 5.95 | 47.50± 12.40 |

## FIGURE CAPTIONS

FIG. 1 — Plot of number of objects vs. absolute $B$ magnitude for the A2199 data. Panel ($a$) shows raw numbers of objects (before subtraction of background and foreground contaminants), whereas panel ($b$) shows the result after subtraction of contaminants. The *solid squares* with error bars represent the data. The *sold line* represents our best maximum likelihood fit to the data ($\alpha \simeq -2.2$). The *dotted line* in panel (a) shows the extimated background (see text for details). The *dashed line* shows what the "Virgo" LF ($\alpha = -1.3$) would look like, forced through the data at the bright end. This illustrates the discrepancy between our data and a "flat" LF.

FIG. 2 — As Fig. 1, for the $I$ band LF derived from the sum of A2052, A2107 and A2666